\documentclass[]{article}
\usepackage[margin=1in]{geometry}
\usepackage[figuresright]{rotating}

\usepackage[utf8]{inputenc}
\usepackage{graphicx}
\usepackage{caption, floatrow}
\usepackage{xcolor, soul}
\usepackage{colortbl}
\usepackage{multirow,multicol}
\usepackage{bigints}
\usepackage{eurosym}
\usepackage{algorithm}
\usepackage[norelsize,ruled,titlenotnumbered,algo2e]{algorithm2e}
\usepackage{authblk}
\usepackage{stfloats}
\usepackage{tabto}
\usepackage{chngcntr}

\usepackage{float}
\usepackage{natbib}
\usepackage{apalike}
\usepackage{subfig}
\usepackage{amsfonts}
\usepackage{amssymb}
\usepackage{amsmath}
\usepackage{array}
\usepackage{numprint}
\usepackage{forest}
\usepackage{subfig}

\newcommand{\argmax}{\operatorname*{\arg\max}}

\usepackage[font=small,labelfont=bf]{caption} % For some reason this package generate a conflic with references
\usepackage{setspace}
\usepackage{sectsty}
\sectionfont{\large\uppercase}
\usepackage{tikz}
\usetikzlibrary{arrows,automata,backgrounds,fit,patterns,petri,shadows,shapes}

\pgfdeclarepatternformonly{stripes}{\pgfpoint{0cm}{0cm}}{\pgfpoint{1cm}{1cm}}{\pgfpoint{1cm}{1cm}}{
  \foreach \i in {0.125, 0.375,...,0.875}{
    \pgfpathmoveto{\pgfpoint{\i cm}{0cm}}
    \pgfpathlineto{\pgfpoint{1cm}{1cm - \i cm}}
    \pgfpathlineto{\pgfpoint{1cm}{1cm - \i cm + 0.125cm}}
    \pgfpathlineto{\pgfpoint{\i cm - 0.125cm}{0cm}}
    \pgfpathclose
    \pgfusepath{fill}
    \pgfpathmoveto{\pgfpoint{0cm}{\i cm}}
    \pgfpathlineto{\pgfpoint{1cm - \i cm}{1cm}}
    \pgfpathlineto{\pgfpoint{1cm - \i cm - 0.125cm}{1cm}}
    \pgfpathlineto{\pgfpoint{0cm}{\i cm + 0.125cm}}
    \pgfpathclose
    \pgfusepath{fill}
  }
}
\tikzstyle{decisionA} = [regular polygon, regular polygon sides = 4, thick, minimum size = 2.7cm, inner sep = 0.1pt, draw = black, fill = gray!40]
\tikzstyle{decisionC} = [regular polygon, regular polygon sides = 4, thick, minimum size = 2.5cm, inner sep = 0.7pt, draw = black]
\tikzstyle{utilityA} = [regular polygon, regular polygon sides = 6, thick, minimum size = 2.5cm, inner sep = 0.1pt, draw = black, fill = gray!40]
\tikzstyle{utilityC} = [regular polygon, regular polygon sides = 6, thick, minimum size = 2.5cm, inner sep = 0.1pt, draw = black]
\tikzstyle{chanceA} = [circle, thick, minimum size = 2cm, inner sep = 0.1pt, draw = black, fill = gray!40]
\tikzstyle{chanceC} = [circle, thick, minimum size = 2cm, inner sep = 0.1pt, draw = black]
\tikzstyle{chance} = [circle, thick, minimum size = 2.3cm, inner sep = 0.1pt, draw = black, pattern = stripes, pattern color = gray!40]
\tikzstyle{empty} = [circle, line width = 0pt, minimum size = 1cm, inner sep = 0.1pt]
\tikzstyle{startstop} = [rectangle, rounded corners,text width=2.5cm, minimum width=2.5cm, minimum height=1cm,text centered, draw=black]
\tikzstyle{process} = [rectangle, minimum width=3.1cm, minimum height=1cm, text centered, text width=3.1cm, draw=black]
\tikzstyle{decisionS} = [diamond, minimum width=1.5cm,text width=1.5cm, minimum height=1cm, text centered, draw=black]
\tikzstyle{decisionM} = [diamond, minimum width=2cm,text width=2cm, minimum height=1cm, text centered, draw=black]
\tikzstyle{arrow} = [thick,->,>=stealth]
\tikzstyle{blockRectangle} = [rectangle,draw = black]
\tikzstyle{block} = [rectangle, draw, fill=white, text width=5.3em, text centered, rounded corners, minimum height=3em]
\tikzstyle{line} = [draw, -latex']

\subsubsectionfont{\normalfont\itshape}

\author{Alberto Redondo}
\author{David Ríos Insua}
\affil{Institute of Mathematical Sciences, ICMAT-CSIC, Spain}

\author{}
\title{Protecting from Malware Obfuscation Attacks through Adversarial Risk Analysis}
\date{}

\begin{document}

\maketitle
\thispagestyle{empty}
\newpage
%\maketitle
%\newpage

\begin{abstract}
Malware constitutes a major global risk affecting millions of users each year. 
Standard algorithms in detection systems perform insufficiently when dealing with 
malware passed through obfuscation tools. We illustrate this studying in detail an open
source metamorphic software, making use of a hybrid framework to obtain the relevant features
from binaries. We then provide an improved alternative solution based on adversarial risk 
analysis which we illustrate describe with an example.

\bigskip
\end{abstract}

\thispagestyle{empty}
\newpage
\setcounter{page}{1}

%%%%%%%%%%%%%%%%%%%%%%%%%%%%%%%%%%%%%%%%%%%%%
\section{Introduction}
\label{sec:intro}
The digital era is bringing along new global threats among which cybersecurity related ones emerge as truly worrisome, see for example the evolution of the Global Risks Map from the \cite{theGlobalRisksReport2017,theGlobalRisksReport2018,theGlobalRisksReport2019}. Indeed, the operation of critical cyber infrastructures relies on components which could be cyber attacked, both incidentally and intentionally, suffering major performance degradation, \cite{rao2016defense}. A key concern is malware (an acronym for malicious software) which, according to \cite{enisa2019}, is among the top threats in the cybersecurity landscape. Indeed, malware \citep{checksummingTechniquesForAntiviralPurposes1992} in its many forms, including trojans, worms, viruses, spyware or adware, affect millions of hosts each year, \cite{malwareBytesReport2018}. Moreover, as reflected in \cite{couce2019}, the negative impacts of such threats may include not only purely financial costs, but also deaths and injuries when dealing with cyber-physical systems, going through stolen personal identifiable information or business secrets in enterprise systems. 

Detection systems are important components in cybersecurity risk management frameworks, see \cite{barrett2018framework}. Anti-malware tools based on scanning file signatures used to recognise most malware until relatively recently. However, these tools are much less effective nowadays due to the continuous changes introduced in such software, as attackers learn how ICT systems owners advance in protection measures. \cite{elingiusti2018malware} and \cite{ye2017survey} provide surveys of current methodologies for malware detection, usually classified in three categories: static, dynamic and hybrid. Static analysis extract relevant binary information from the software without running it; dynamic methods are carried out in separate isolated environments, like sandboxes, to extract relevant information from the running software; hybrid methods combine both approaches, typically allowing us to gain better information and understanding of the behaviour of the incumbent binary file. Examples include \cite{santos2009n}, who introduce a static analysis with Portable Executable (PE) files and describe that malware detection works reasonably well based on 
operational codes (OpCodes); \cite{anderson2012improving} who combine both approaches through kernel analysis; and \cite{van2014sssm} who use semantic sets and N-gram bytes with a Naive Bayes (NB) classifier jointly with an API function-based signature. Our benchmark will be \cite{o2016detecting} as their binary dataset was available and we could use it for comparison. They use OpCode density histograms extracted during run time execution and classify with Support Vector Machines (SVM).

\begin{figure}[H]
\centering
\begin{tikzpicture}[node distance = 3cm, auto]
    % Place nodes
    \node [block] (block1) {\small No Stealth \\ 1970};
    \node [block, right of=block1, node distance=2.8cm] (block2) {\small Encrypted    \\ 1987};
    \node [block, right of=block2, node distance=2.8cm] (block3) {\small Oligomorphic \\ 1990};
    \node [block, right of=block3, node distance=2.8cm] (block4) {\small Polymorphic  \\ 1990};
    \node [block, right of=block4, node distance=2.8cm] (block5) {\small Metamorphic  \\ 1998 / now};
    % Draw edges
    \path [line] (block1) -- (block2);
    \path [line] (block2) -- (block3);
    \path [line] (block3) -- (block4);
    \path [line] (block4) -- (block5);
\end{tikzpicture}
\caption{The malware camouflage progression, \cite{rad2012camouflage}}
\label{fig:camouflageMalwareProgression}
\end{figure}
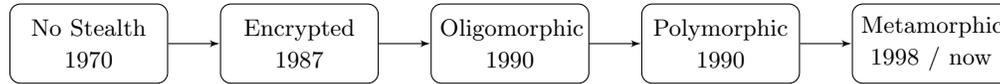

A prominent attacking strategy through malware is obfuscation which designates a group of procedures that make a malware binary more difficult to be detected through anti-malware tools, as reflected in the camouflage malware progression presented in Figure \ref{fig:camouflageMalwareProgression}. The term stealth is used when a binary hides its code to other programs; this method was not considered effective as antimalware tools were able to find the benign parts of the code and detect the remaining malware portion. The next obfuscation advance was based on encryption, so that the code included a loop that encrypts its body. Then, attackers tried oligomorphic malware which includes loops using predefined forms for each malware copy; however, once antimalware tools were redesigned to search for all predefined loop combinations, malware became vulnerable again.
\noindent Currently, obfuscation techniques have become very 
sophisticated and are based on two principles:  poly and metamorphism, \cite{you2010malwareObfuscationTechniques}. Polymorphic methods encrypt the malware body, changing its form for each malware copy. Metamorphic methods use advanced obfuscation based on employing mutation engines which modify the whole binary: consequently, each new malware clone has a different code sequence, size, structure or syntactic properties making highly difficult for anti-malware tools to detect it. 

A few approaches have been used to detect obfuscation attacks. \cite{lakhotia2005method} obtain stack operations to detect obfuscated functions, but found that the stack was easy to corrupt. \cite{rolles2009unpacking} proposes a circumvention method to break a virtualisation obfuscator converting the code to byte code language; after conversion, he applies reverse engineering techniques. \cite{kakisim2018analysis} provide several algorithms to detect metamorphic malware based on hidden Markov models, $K$-means clustering, artificial neural networks, Bayesian networks and decision trees. \cite{kaushal2012metamorphic} count the API call frequency to detect metamorphic malware which performs useless code insertion, register usage exchange, code reordering through jump instructions or equivalent instruction replacement.

In this work, we propose a methodology to protect from obfuscation attacks based on Adversarial Risk Analysis (ARA), \cite{banks2015adversarial}. Section 2 presents a framework for malware detection based on hybrid analysis which provides our initial benchmark. Section 3 illustrates the problems entailed by metamorphic malware rendering standard methods less effective. We then detail our ARA model to detect obfuscation attacks over the benchmark performed. We finally provide examples and conclude with a discussion.

%%%%%%%%%%%%%%%%%%%%%%%%%%%%%%%%%%%%%%%%%%%%%
\section{A Framework to Protect from Malware Using a Hybrid Approach}
\label{sec:framework}
\begin{figure}[H]
\centering
    \includegraphics[width=1 \textwidth]{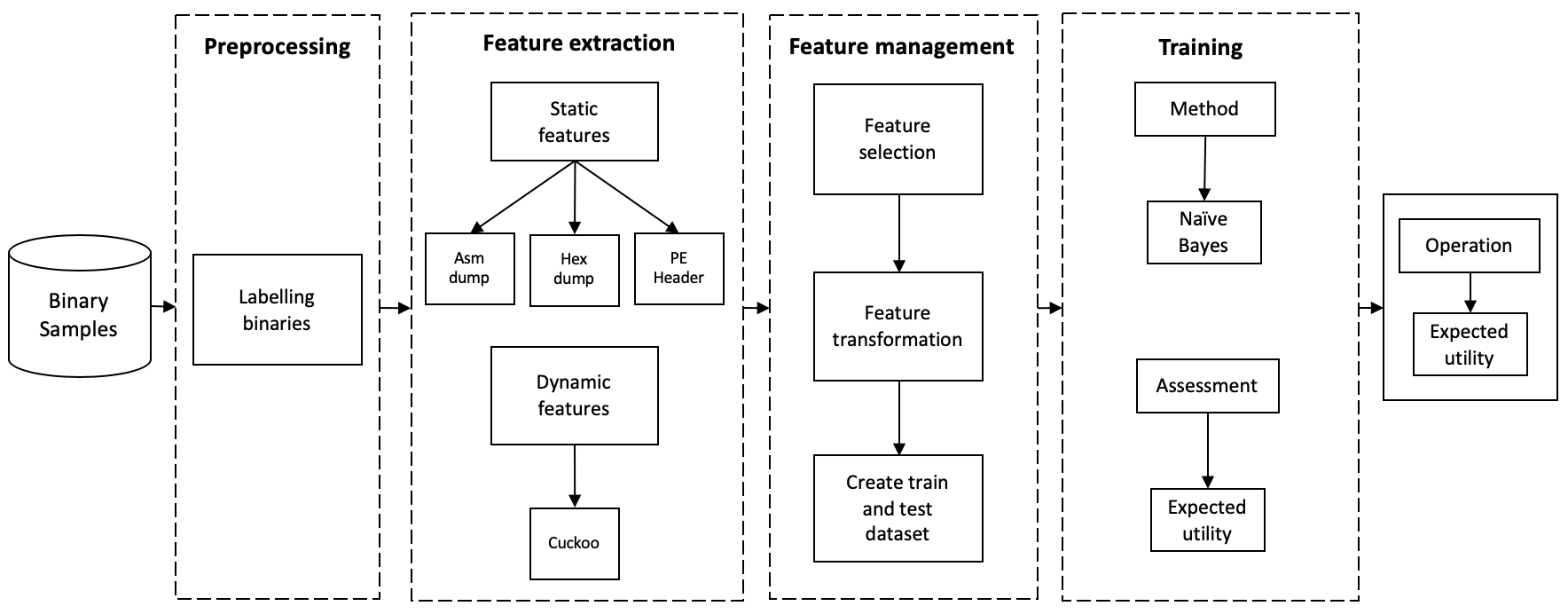}
\caption{Malware detection framework}
\label{fig:MalwareDetectionFramework}   
\end{figure}

The approach initially proposed to detect malware uses a combination of static and dynamic methods and comprises five stages, Figure \ref{fig:MalwareDetectionFramework}: preprocessing, feature extraction, feature management, training and operation. It will serve as benchmark for our later developments.
%%%%%%%%%%%%%%%%%%%%%%%%%%%%%%
\subsection{Preprocessing}
\label{subsec:prepro}
We start with a set of binary samples to train a classifier. Each binary is labelled as malware (M) or benign (B). If some of the binaries available are not labelled, we assess their label through a tool called VirusTotal (VT), \cite{virusTotal}. VT uses $70$ online anti-malware systems; we consider a binary as malware when at least $50\%$ of them indicate so.
%%%%%%%%%%%%%%%%%%%%%%
\subsection{Feature extraction}
\label{subsec:featureExtraction}
Feature extraction combines the procurement of static features directly from the binary and dynamic features when executed in a controlled environment. For our hybrid approach, we use the static Assembly Language File (ASM) features considered by \cite{derekChadwick} as well as the Hexadecimal dump (Hex dump) and the Portable Executable File Header (PE Header) static features  determined by \cite{ahuja2017robust}, who claim to have achieved good results in malware detection.
%%%%%%%%%%%%%%%%%%%%%
\subsubsection{Static features}
\label{subsubsec:staic}
They are divided into three categories depending on the method employed to extract them: 
\begin{itemize}
\item ASM. We convert the binaries into assembled files through objdump, \cite{objectdump}. Then, we extract its sections, registers, OpCodes, API calls and keywords.
\item Hex dump. We transform the binaries into the hexadecimal format through hexdump, \cite{hexdump}. We then obtain the file size, the mean, median, maximum and minimum of the binary entropy, along with the entropy of the whole binary and the entropy variance and range.
\item PE Header, \cite{wang2009detecting}. The standard file format allows us to extract features from the binaries such as the size of the code and the number of sections, symbols and imports.
\end{itemize}   
Table \ref{tab:featureExtractionTypes} summarises the $1068$ static features extracted. 
\begin{table}[H]
\footnotesize
  \centering
  \begin{tabular}{ | c | c | c | c |}
    \hline
    {\bf Feature extraction method} & {\bf Features}      & {\bf Type}        &  {\bf Total}    \\ \hline
    ASM                       & Sections    & Discrete    &   $9$     \\ \hline
    ASM                       & Registers   & Discrete    &   $26$    \\ \hline
    ASM                       & Opcodes     & Discrete    &   $94$    \\ \hline
    ASM                       & API calls   & Discrete    &   $794$   \\ \hline
    ASM                       & Keywords    & Discrete    &   $95$    \\ \hline
    Hex dump                  & Entropy     & Continuous  &   $7$     \\ \hline
    Hex dump                  & File size   & Discrete    &   $1$     \\ \hline
    PE Header                 & Several     & Discrete    &   $42$    \\ \hline
  \end{tabular}
\caption{Static features extracted}\label{tab:featureExtractionTypes}
\end{table}
%%%%%%%%%%%%%%%%%%%%%%%%%%%%%%%%%%%%
\subsubsection{Dynamic features}
Dynamic features are generated based on the run time behaviour of the binaries executed within a Virtual Machine. To perform such analysis we use the \cite{cuckoo} environment, which generates a report from the behaviour obtained over a fixed period of time. We use a two minutes default configuration. We obtain the reports from all the binaries selecting twelve features which we consider relevant for malware detection including the number of mutex (used to coordinate program processes when the program has to perform several tasks simultaneously), the number of file operations such as the files read or deleted (used by some malware to steal or remove information to cause fraud or damage), the register operations and the dll libraries loaded.
%%%%%%%%%%%%%%%%%%%%%%%%%%%%%%%%%%%%
\subsection{Feature management}
\label{subsec:featureManagement}
In this phase, we perform feature management. We select the most suitable features for the incumbent classifier in the problem at hand and study the data types which may be transformed into categorical, discrete or continuous, looking for those providing best results. This may entail repeated iterations through this step and \ref{subsec:training}. We then randomly split the dataset into train and test subsets, according to a division ratio e.g. of $0.20$: $80\%$ of the set is used to train and the remaining $20\%$ for testing purposes.
\subsection{Training}
\label{subsec:training}
At this stage, we must choose the classification algorithms to be used for malware detection. As the quantity of data in the digital era is enormous, for instance, it is typical that threat intelligent systems have to process hundreds of thousands of binaries per day to determine whether they are malware or not, we need classifiers that obtain results in a reasonable time, see \cite{ye2017survey} for a survey. 

In this paper, we consider as benchmark a NB classifier, \citep{lewis1998naive}. This classifier has been used for malware detection by \cite{sahay2019efficient}, among many other authors. NB requires short computational time for training compared with others and also, it is highly scalable, see \cite{ashari2013performance}. We shall train the NB classifier based on maximising expected 
utility. A typical utility in this context is the $0-1$ utility (1 if it classifies correctly; 0, otherwise), implicitly leading to standard classification criteria based on detection accuracy (DA), and the false positive (FPR) and false negative (FNR) rates. However, in a more risk analytic fashion, we could consider other utility functions, see our discussion in Section \ref{subsec:otherUtilites}.
%We must select the algorithms to be used for malware detection, see \cite{ye2017survey} for a survey. In this paper, we consider as benchmark on NB classifier, \citep{lewis1998naive}. NB has been used for malware detection by, among many others, \cite{sahay2019efficient}, as the quantity of data in the digital era is enormous, it is necessary to use classifiers that obtain results in a reasonable time. For example, it is typical that threat intelligent systems have to process hundreds of thousands of files per day to determine whether they are malware or not. Compared with other classifiers, NB requires short computational time for training and also, it is highly scalable, see \cite{ashari2013performance} for NB classification time comparisons, we shall train the classifier based on maximising expected utility. A typical utility in this context is the $0$-$1$ utility ($1$ if you classify properly; $0$, otherwise). This utility leads to the standard classification criteria like the detection accuracy (DA) and the false positive (FPR) and false negative (FNR) rates.
%%%%%%%%%%%%%%%%%%%%%%%%%%%%%%%%%%%%%%%%
\subsection{Operation}
\label{subsec:operation}
Once we have properly chosen the features and trained our classifier, we set up the framework into operation. When we receive a new binary, we extract its features and process them through the classifier to decide whether it corresponds to benign software or to malware. The operational criteria used, as in standard risk analysis \cite{bedford2001probabilistic}, is based on maximising expected utility through
\begin{equation}
c(x)=\argmax_{y_{C}} \sum_{y} u_{C}(y_C,y)p(y|x),
\label{eq:operation}
\end{equation}
where $c(x)$ is the class (malware, benign) which maximises expected utility, $y_{C}$ is the class to be predicted for the binary, $u_{C}$ is the utility function used by the organisation and $p(y|x)$ is the probability of the binary belonging to class $y$ given the features $x$.
%%%%%%%%%%%%%%%%%%%%%%%%%%%%%%%%%%%%%%%%%%%
\subsection{Benchmark experiments}
\label{subsec:benchmarkExperiments} % 
The experiments \footnote{For reproducibility purposes, code is available at https://github.com/aRedondoH/MalwareDetectionCluster.git and at https://github.com/aRedondoH/AROA.git.} reported here were performed in a distributed system with $25$ nodes (16 cores, each at 2.60GHz). The hybrid framework has been implemented in Python $3.6.7$ with packages such as $sklearn$, $numpy$, $pandas$ and $joblib$, and it has been tested with malware datasets from \cite{heavenvx} and Virus Total, \cite{virusTotal}. Experiments refer to two different years to showcase and understand possible deterioration over time due to novel attacking techniques. The datasets respectively consist of $2698$ and $2955$ malware binaries, which we merge with $542$ benign binaries obtained from clean copies of MS Windows 7 and 8 from the Program Files folder.

To evaluate the framework performance initially, we create an experiment based on: building a dataset with $542$ malware binaries randomly selected, merged with $542$ benign binaries; extract the features of the binaries, as described in Section \ref{subsec:featureExtraction}; split the obtained dataset into training and test sets; and, train the NB classifier with the training set. Then, we undertake the NB based classification through the test set. This experiment is repeated $1000$ times with the VxHeaven and Virus Total sets. The results are shown in Table \ref{tab:averageVxVt}, where we include the results from \cite{o2016detecting} for comparison.
\begin{table}[H]
\centering
\scalebox{0.83}{
\begin{tabular}{c|c|c|c|c|c|c|}
\cline{2-7}
& \multicolumn{3}{|c|}{\bf VxHeaven} & \multicolumn{3}{|c|}{\bf Virus Total}\\
\cline{1-7}
\multicolumn{1}{|c|}{\bf Method} & {\bf DA}   & {\bf FPR}  & {\bf FNR} &  {\bf DA}  &  {\bf FPR}   & {\bf FNR} \\
\hline
\multicolumn{1}{|c|}{O'Kane et al. (SVM)}  & 0.86   & -  & 0.15 &  -  &  -   &  -  \\
\hline
\multicolumn{1}{|c|}{NB}  & 0.93   & 0.11  & 0.02 &  0.86  &  0.21   &  0.03  \\
\hline
\end{tabular}
}
\caption{Averaged detection accuracy for VxHeaven and Virus Total datasets}\label{tab:averageVxVt}
\end{table}
\noindent Our benchmark achieved $86\%$ DA with the VxHeaven dataset. With our feature selection process, we improve their performance by a $7\%$ DA using NB. Note though that with the more modern Virus Total dataset, the classifier decreases by $8\%$ its DA (with an increment of $10\%$ in the FPR). Recall that the time difference between both datasets was of about five years. Over that period, programming languages have evolved, programming styles have changed and new obfuscation tools have appeared. This could explain such performance degradation.

%%%%%%%%%%%%%%%%%%%%%%%%%%%%%%%%%%%%%%%%%%%%%
\section{The impact of obfuscated malware}
\label{sec:impactObfuscatedMalware}
As we have seen, we may achieve reasonably good detection results through the proposed hybrid approach, accomplishing better performance than e.g. \cite{o2016detecting} benchmark. However, as mentioned in Section \ref{sec:intro}, an attacker could create obfuscated malware, critically affecting the performance of detection algorithms. This is the general issue addressed in the emergent field of Adversarial Machine Learning, see the recent reviews in \cite{vorobeychik2018adversarial} and \cite{joseph2019adversarial}. 

\begin{figure}[H]
\centering
\begin{tikzpicture}[->, >=stealth',shorten >= 1pt, auto,semithick,scale=0.45, node distance = 2.5cm]
    \node[blockRectangle](binNode)[]{\footnotesize Binary};
    \node[text width=2.1cm,blockRectangle](staticNode)[right of = binNode, yshift=0.8cm, xshift=0.4cm]{\footnotesize$(x_{11}, \dots, x_{16})$,\\$(x_{17},\dots,x_{1m_{1}})$};
    \node[empty](labelStaticFeatures)[above of = staticNode,yshift=-1.7cm, xshift=-0.1cm]{\footnotesize Static features};
    \node[text width=2.1cm, blockRectangle](dynamicNode)[below of = staticNode,yshift=0.9cm]{\footnotesize $(x_{21}, \dots, x_{2m_{2}})$};
    \node[empty](labelDynamicFeatures)[below of = dynamicNode,yshift=1.9cm, xshift=0cm]{\footnotesize Dynamic features};
    \node[blockRectangle](metameNode)[right of = dynamicNode, yshift=0.9cm, xshift=0.2cm]{\footnotesize Metame};
    \node[text width=2.1cm,blockRectangle](staticNodeObfu)[right of = metameNode,yshift=0.7cm, xshift=0.4cm]{\footnotesize $(x'_{11}, \dots, x'_{16})$,\\$(x_{17},\dots,x_{1m_{1}})$};
    \node[text width=2.1cm,blockRectangle](dynamicNode2)[right of = metameNode,yshift=-0.9cm, xshift=0.4cm]{\footnotesize $(x_{21},\dots,x_{2m_{2}})$};
    \path
    (binNode) edge[out = 0 , in = 180] node{}(staticNode)
    (binNode) edge[out = 0 , in = 180] node{}(dynamicNode)
    (staticNode) edge[out = 0 , in = 180] node{}(metameNode)
    (dynamicNode) edge[out = 0 , in = 180] node{}(metameNode)
    (metameNode) edge[out = 0 , in = 180] node{}(staticNodeObfu)
    (metameNode) edge[out = 0 , in = 180] node{}(dynamicNode2)
    ;
\end{tikzpicture}
\caption{Metame feature modification}
     \label{fig:metameMod}
\end{figure}
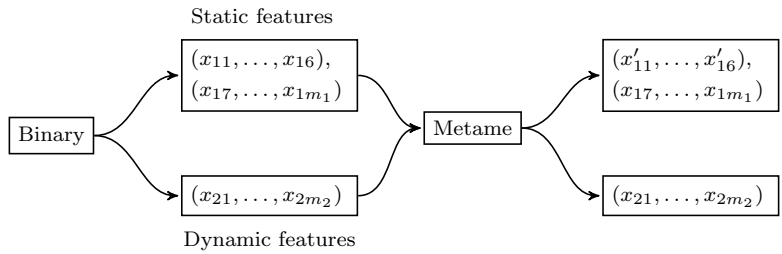

\begin{figure}[H] 
\includegraphics[width=0.60\textwidth]{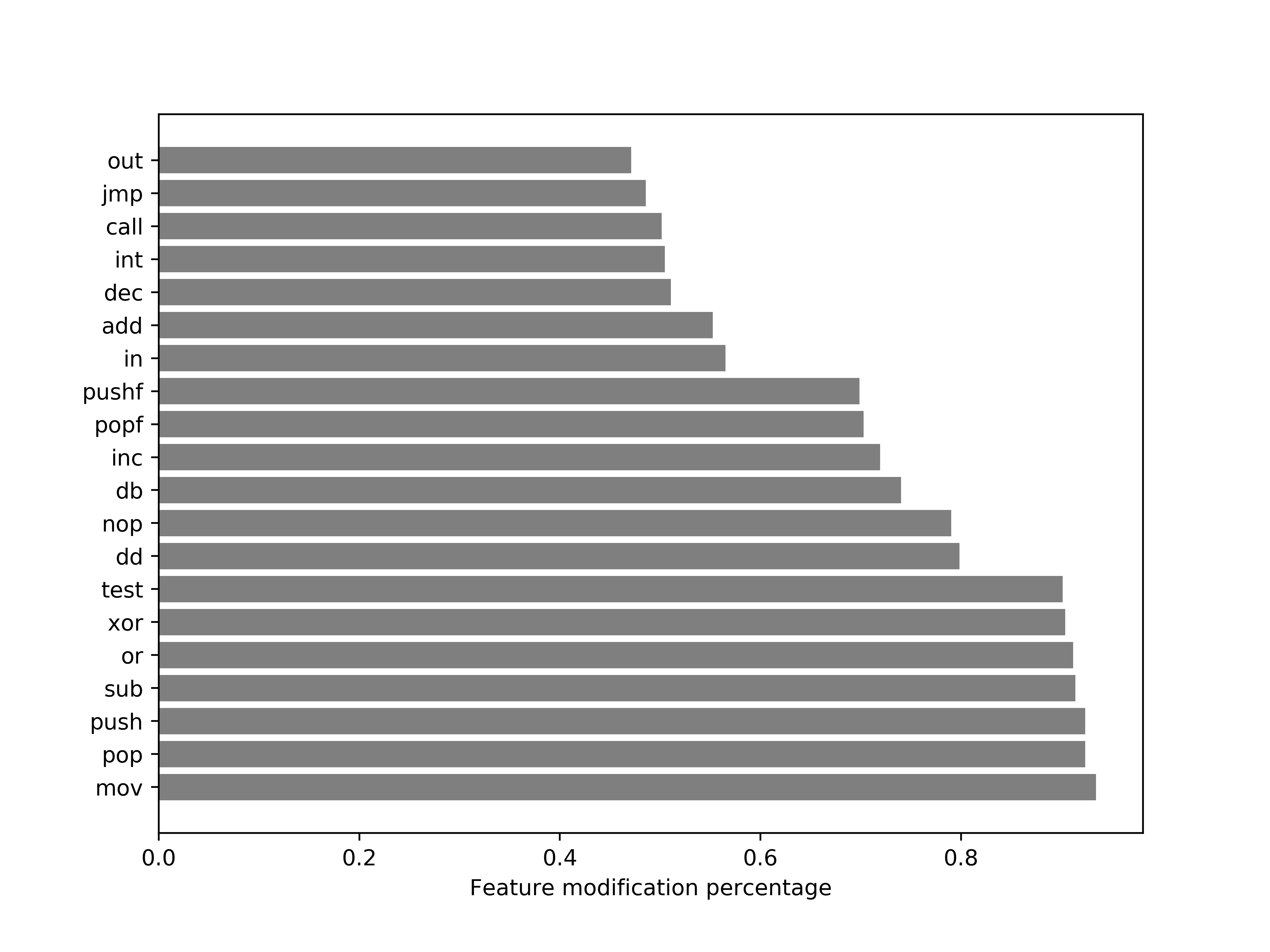}
\caption{Top 20 features affected by metame ranked from most modified to least one}
\label{fig:top20FeaturesAffected}
\end{figure}

\begin{figure}[H] 
    \centering
    \subfloat[Feature convergence]{
        \includegraphics[height=5cm,width=0.47\textwidth]{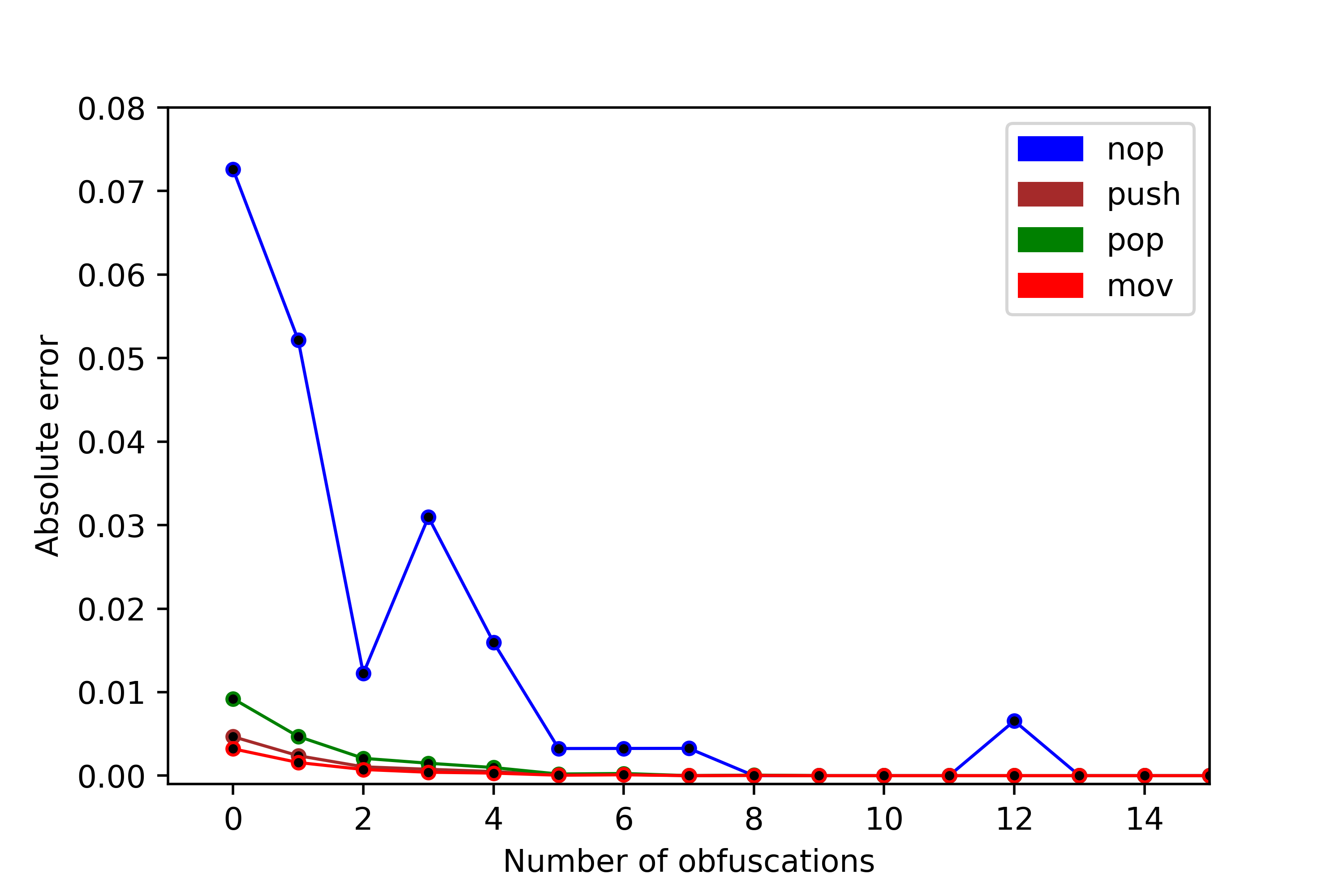}
        \label{fig:metameConvergence}
        }
    \hfill
    \subfloat[Feature divergence]{
        \includegraphics[height=5cm,width=0.47\textwidth]{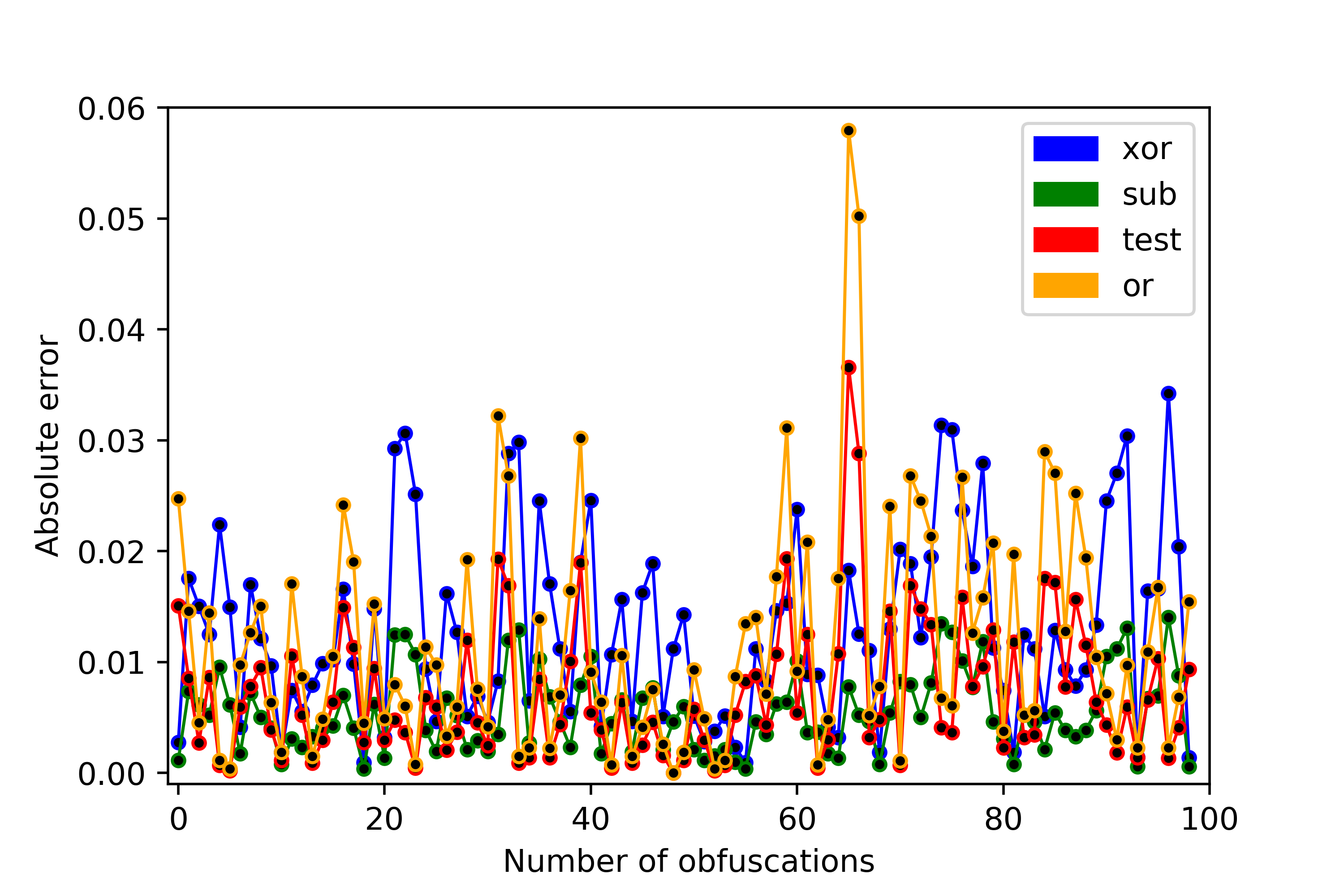}
        \label{fig:metameDivergence}
        }
    \caption{Evolution of features after repeated obfuscations}
\end{figure}

To illustrate and understand this point, we shall obfuscate malware through an open source metamorphic software obfuscation engine called metame, \cite{metame}. For a given code, this tool creates a software clone with the same behaviour, but different structure, aimed at overcoming detection barriers. For this, metame disassembles the binary seeking for six types of OpCodes, designated $nop$, $xor$, $sub$, $push$, $pop$, and $or$, and performs operations with them. For instance, some $xor$ OpCodes are replaced by $sub$ ones. This process modifies only static features, not dynamic ones, as metame aims at preserving the original software behaviour. Figure \ref{fig:metameMod} reflects this process, where we emphasize the six features that metame modifies.  
\noindent As a consequence, during the obfuscation process, the number of $xor$ OpCodes will decrease, whereas that of $sub$ OpCodes would increase. 
%As an example, Table \ref{tab:binObfuscated} presents the results of obfuscating a specific binary showing how the corresponding features changed. 
%\begin{table}[H]
%\begin{tabular}{| l | l | l | l | l | l | l | l |}
%\hline
%{\bf Binary}  & {\bf nop} & {\bf xor} & {\bf sub} & {\bf push} & {\bf pop} & {\bf or} \\ \hline
%Before &    20     &    184    &    325    &    1005    &    181    &    49   \\ \hline
%After    &    66     &    160    &    360    &    1154    &    330    &   105   \\ \hline
%\end{tabular}
%\caption{Feature changes in an obfuscated binary}\label{tab:binObfuscated}
%\end{table}
Note that due to these modifications other features could be affected indirectly. To observe
such changes, we obfuscated $3000$ binaries counting the instances in which each feature was modified through obfuscation. Figure \ref{fig:top20FeaturesAffected} shows the $20$ features most modified, ranked according to the number of times they were modified.
\noindent To further understand the obfuscation process, we applied it $100$ times to a specific binary to check whether there were features that might stop changing after a certain number of obfuscations. Figure \ref{fig:metameConvergence} represents the evolution of the modification percentage of the $nop$, $push$, $pop$ and $mov$ features after $14$ obfuscations, whereas Figure \ref{fig:metameDivergence} illustrates that of the $xor$, $sub$, $test$ and $or$ features after $100$ obfuscations.
\noindent We notice that the first features practically stop changing after $13$ obfuscations. However, the features reflected in Figure \ref{fig:metameDivergence} keep on varying their values throughout the repeated application of the obfuscator.

\begin{figure}[H] 
\includegraphics[width=0.47\textwidth]{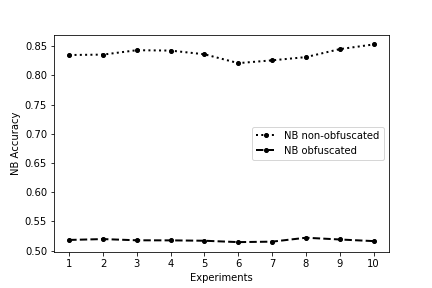}
\caption{Obfuscation degrades malware detection}
\label{fig:nbNoObfuVsObfu}
\end{figure}

To assess the impact of obfuscation on the detection process, we trained a NB classifier with non-obfuscated VT (2018) malware mixed with benign binaries ($50\%$ malware, $50\%$ benign) as in in Section \ref{subsec:benchmarkExperiments}. Then, we predicted through a test set formed by metame obfuscated VT malware mixed with benign binaries, with results in Figure \ref{fig:nbNoObfuVsObfu}, where we show the detection accuracy with non-obfuscated and obfuscated malware over $1000$ experiments, with results averaged over groups of $100$ experiments. Observe that the accuracy of the classifier degrades approximately by $31\%$ showing the potential of obfuscation methods in fooling standard classifiers.

\begin{figure}[H] 
\includegraphics[width=0.47\textwidth]{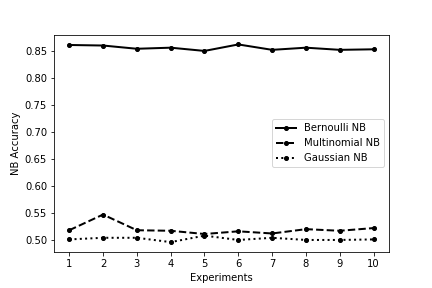}
\caption{Feature management might improve performance}
\label{fig:nbComparison}
\end{figure}

Given this performance degradation, we explored whether certain feature management operations may improve results somehow robustifying algorithms. For that, as described in Section \ref{subsec:featureManagement}, we transformed the data into different types trying a Bernoulli NB with binary features (expressed with $1$ when the feature is present and $0$ when it is absent); a Multinomial NB with discrete features (representing the feature frequency in the binaries), \cite{singh2019comparison}; and a Gaussian NB with continuous features (assuming they are distributed according to a normal distribution), \cite{yilmaz2019locally}, to determine the classifier providing better results. As shown in Figure \ref{fig:nbComparison}, the Bernoulli NB classifier performs better because of its underlying modelling assumptions.

\begin{figure}[H] 
\centering
\includegraphics[width=1\textwidth]{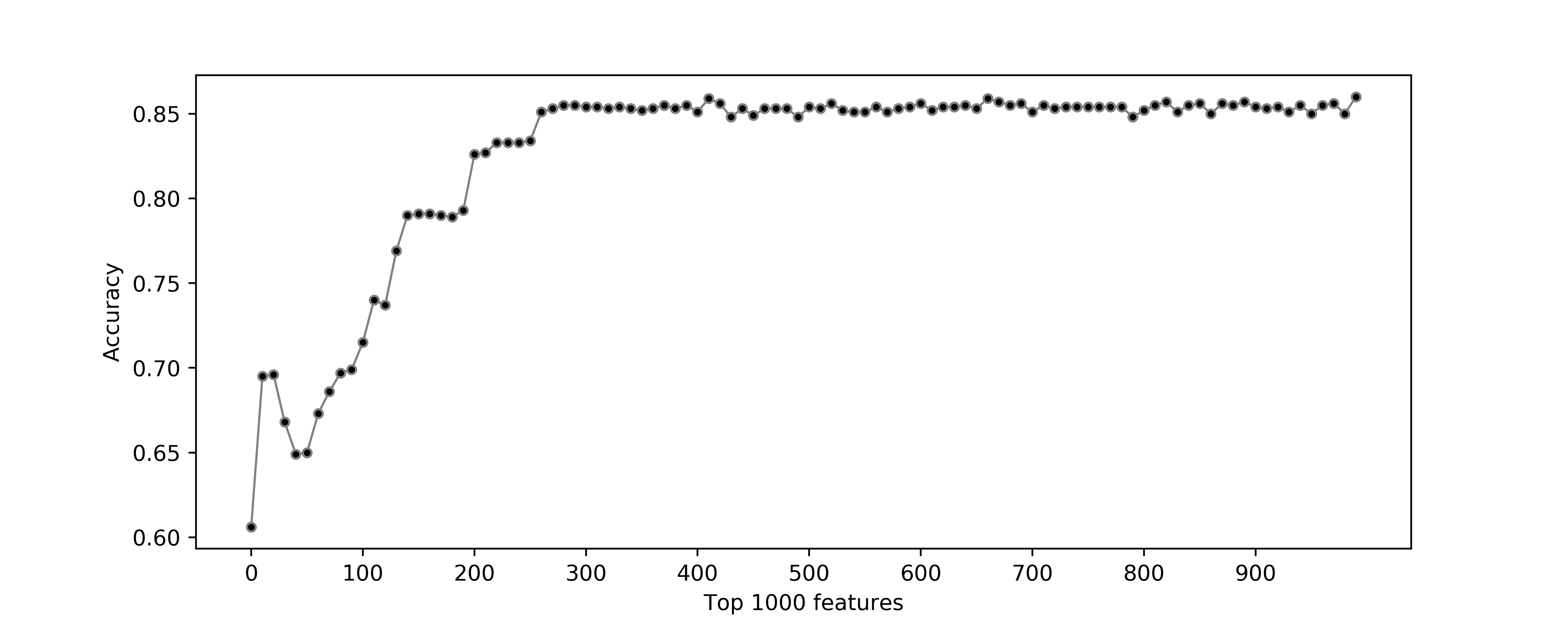}
\caption{Performance with top 1000 features}
\label{fig:top1000FeaturesStudy}
\end{figure}

After such feature transformation, we also checked for the minimum number of features needed to reach a reasonable accuracy. This is important so as to reduce computation times in presence of very large amounts of binaries. We performed experiments adding features one by one to the classifier to observe the accuracy evaluation, Figure \ref{fig:top1000FeaturesStudy}, based on the ranking in Figure \ref{fig:top20FeaturesAffected} (extended to the $1068$ features).
\noindent We observe that with about $120$ features, NB achieves around $80\%$ accuracy and with around $220$ features, it reaches around $85\%$.

%%%%%%%%%%%%%%%%%%%%%%%%%%%%%%%%%%%%%%%%%%%%%
\section{Adversarial Risk Analysis Against Obfuscation Attacks}
\label{sec:aroa}
We have discussed how a classifier which operates under the standard maximum expected utility risk analytic criteria in (\ref{eq:operation}) may see its performance deteriorated if we do not
pay attention to the presence of attackers ready to modify the features of a malware. Thus, we need to take into account the presence of such adversaries when making the detection decisions. The model that we shall propose to detect obfuscation attacks will adapt the adversarial risk
analysis approach in \cite{naveiro2019adversarial}. We sketch the common elements with that framework, emphasising the differences proposed to detect malware. 

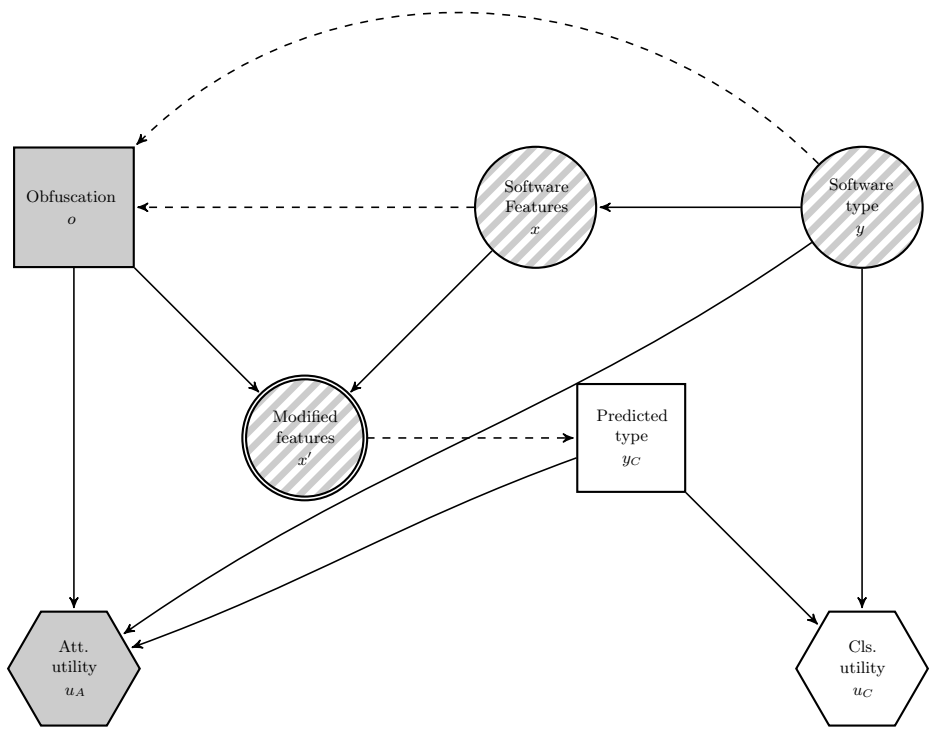
\begin{figure}[H]
\centering
\begin{tikzpicture}[->, >=stealth', shorten >= 1pt, auto, node distance = 6.2cm, semithick, scale = 0.7, transform shape]
\begin{scope}[xscale=-1]
\node[text width=1.4cm,decisionC ](C) []{\small \reflectbox{Predicted}\\$\,\,\,\,\,\,\,$\reflectbox{type}\\ $\,\,\,\,\,\,\,$ \reflectbox{$y_C$}};

\node[text width=1.7cm,chance,accepting](XS)[right of = C]{\small $\,\,\,\,$\reflectbox{Modified} \\ $\,\,\,$ \reflectbox{features}\\ $\,\,\,\,\,\,\,\,\,\,\,\,\,$\reflectbox{$x'$}};

\node[text width=1.3cm,chance](Y) [above left of = C]{\small $\,$\reflectbox{Software} \\ $\,\,\,\,$ \reflectbox{type} \\ $\,\,\,\,\,\,\,\,\,\,\,$\reflectbox{$y$}};

\node[text width=1.8cm,chance](X) [above left of = XS]{\small $\,\,\,\,\,$\reflectbox{Software} \\ $\,\,\,$ \reflectbox{Features}\\$\,\,\,\,\,\,\,\,\,\,\,\,\,\,\,$\reflectbox{$x$}};

\node[text width=1.6cm,decisionA](A) [above right of = XS]{\small \reflectbox{Obfuscation}\\ $\,\,\,\,\,\,\,\,\,\,\,\,$ \reflectbox{$o$}};
 
\node[text width=1cm,utilityC](UC)[below left of = C ]{\small  $\,\,\,\,$\reflectbox{Cls.}\\ $\,$\reflectbox{utility}\\ $\,\,\,\,\,$\reflectbox{$u_C$}};
 
\node[text width=1cm,utilityA](UA)[below right of = XS ]{\small  $\,\,\,\,$\reflectbox{Att.}\\$\,$\reflectbox{utility}\\$\,\,\,\,\,$\reflectbox{$u_A$}};

 \path
 (Y) edge[out =   0, in =  180        ] node {}(X)
 (Y) edge[out =   -90, in = 90        ] node {}(UC)
 (Y) edge[out =   -35, in =  145] node {}(UA)
 (Y) edge[out = 45, in = 135, dashed] node {}(A)
 (X) edge[out =   -45, in = 135, ] node {}(XS)
 (X) edge[out =   0, in =  -180, dashed] node {}(A)
 (A) edge[out =  -135 , in = 45, ] node {}(XS)
 (A) edge[out =   -90, in =  90, ] node {}(UA)
 (C) edge[out =   -135, in =  45, ] node {}(UC)
 (C) edge[out =   -20, in =  160, ] node {}(UA)
 (XS) edge[out =   180, in =  0, dashed] node {}(C)
 ;
 \end{scope}
 \end{tikzpicture}
\caption{Bi-agent influence diagram for the obfuscated malware problem}
\label{fig:biagentProblem}
\end{figure}

We consider a classifier ({\it C}, she="Cleo") aiming at maximising her expected utility when classifying binaries between benign ({\it y=B}) and malware ({\it y=M}). An adversary ({\it A}, he="Alan") is willing to obfuscate binaries maximising his expected utility: he modifies the features $x$ in malware binaries to $x'=o(x)$ through obfuscation to outguess the classifier. We designate such obfuscation as $o_{x \rightarrow x'}$. The problem faced by both agents is represented in Figure \ref{fig:biagentProblem} through a bi-agent influence diagram, \cite{banks2015adversarial}.

\noindent Grey nodes represent issues that affect solely the obfuscator's decisions; white ones, those that impact solely those of the classifier; finally, striped nodes affect both agents'
decisions.  Alan's (the obfuscation attack chosen) is represented through node $o$; Cleo's decision (the classification choice) through $y_{C}$. The impact of obfuscation over $x$ results in the modified features $x'$. The agent's utilities are shown through nodes $u_{A}$ and $u_{C}$, respectively. The classifier needs to determine the class $y$ upon observing $x'$, with her guess $y_{C}$ providing her an utility $u_{C}(y_{C},y)$. Depending on such guess $y_{C}$, the attacker would obtain his utility $u_{A}(y_{C},y,o)$. We consider that the attacker is only interested in obfuscating malware, considering that he makes attacks only in the operational phase. Thus, we assume that training is undertaken with non-obfuscated malware. Therefore, the classifier may estimate the distribution ($p_{C}(y)$) of malware and benign software, as well as the distribution ($p_{C}(x|y)$) of the features given the software type, in our case through NB during the training phase.

%%%%%%%%%%%%%%%%%%%%%%%%%%%%%%%%%%%%%%
\subsection{Classifier's problem}
She deals with the problem as a game from a Bayesian perspective, \cite{kadane1982subjective}, taking into account that Alan's obfuscation decision is random to her. For her analysis, the classifier needs: $p_{C}(y)$, assessing her conviction about the label $y$ of the binary, with $p_{C}(M)+p_{C}(B)=1$ and $p_{C}(M),p_{C}(B) \geq 0$; $p_{C}(x|y)$, which models her beliefs about the features $x$ of the binary, depending on whether it is benign or malware; $p_{C}(x'|o,x)$, describing her beliefs about the impact through an obfuscation attack $o$ over a binary with features $x$; $u_{C}(y_{C},y)$, models her preferences when she predicts class $y_{C}$ for the binary and the actual class is $y$; and, $p_{C}(o|x,y)$, which represents her beliefs over the Attacker's obfuscation action $o$ given a binary with features $x$ and label $y$.

For a given binary characterised by its features $x$, we assume we may consider a set of 
reasonable obfuscations $O(x)$. As a consequence, for a given binary received with features $x'$, we may consider the corresponding $\mathcal{X}'$ of features $x$ that lead to $x'$, when the optimal binary is obfuscated. Specifically, given a binary $x=(x_1,x_2,\dots,x_n)$ with features $x_i \in \{0,1\}$ we designate by $I(x)$ the set of indices with $x_i=0$. Then, the set of possible attacks over the binaries $O(x)=\{O_{H} : H \subset I(x)\}$ where $O_{H}$ represents that the features whose index $j \in H$ are converted into a $1$. Similarly, given a binary $x'$, let $J(x')$ be the set of indices with $x_{i}'=1$. Then, the set of originating binaries is $\mathcal{X}'=\{x_{K} : J(x')\}$, where $x_{K}$ is such that $x_{i}=0$ if $i \in K$, and $x_{i}=x_{i}'$, otherwise. However, to constraint the size of these sets so as to mitigate computational bottlenecks, we introduce a maximum number of originating binaries, randomly sampled, and a maximum number of attacks, again randomly sampled.

Then, given $x'$, Cleo aims at finding the class $c(x')$ maximising her expected utility, which based on the developments in \cite{naveiro2019adversarial}, can be computed (rather than through (\ref{eq:operation}) which would be adversary unaware) through
\begin{eqnarray}
 c(x')  &=& \argmax_{y_C} \bigg[ u_C(y_C, M) p_C(M) \sum_{x \in \mathcal{X}'} p_C(o_{x \rightarrow x'} |x,M) p_C(x|M) \nonumber \\
  &+& u_C(y_C, B) p_C(B)p_C(x'|B)\bigg], \label{pis}
\end{eqnarray} \normalsize
where $p_C(o_{x \rightarrow x'} |x,M)$ models the probability, according to her, that the attacker adopts the obfuscation $o_{x \rightarrow x'}$ %which transforms features $x$ to $x'$
when the malware has original features $x$.

Note that all of the required ingredients in (2) are standard risk analytic assessments, except for $p_{C}(o_{x \rightarrow x'}|x,M)$ due to its strategic component. We facilitate its estimation considering next the problem faced by the Attacker.

%%%%%%%%%%%%%%%%%%%%%%%%%%%%%%%%%%
\subsection{Attacker's problem}
To solve his decision making model we need, in principle: $p_{A}(x'|o,x)$, which assess his beliefs about the obfuscation attacks performed to $x$; $u_{A}(y_{C},y,o)$, describing his preferences when the classifier predicts the label to be $y_{C}$, the actual label is $y$ and the attack is $o$; and, $p_{A}(c(x')|x')$, which expresses his thoughts about the classifier's prediction when she observes the features $x'$ of the (obfuscated) binary. Let $p=p_{A}(c(o(x)) = M|x')$
be the probability that the attacker concedes to Cleo saying that the binary is malware, when she observes $x'$. Since he will have uncertainty about it, we denote its density by $f_{A}(p|o(x))$, and designate its expectation $p_{o(x)}^{A}$. Then, he seeks to maximise his expected utility through
\begin{eqnarray*}
o^*(x, M) &=& \argmax_{o} \int \bigg[ u_A( M, M, o) \, \, p  + u_A( B, M, o) \, \, (1 - p) \bigg] f_A(p|o(x)) dp \\
&=& \argmax_{o} \left[u_A( M, M, o) - u_A( B, M, o)\right] p_{o(x)}^A  +  u_A( B, M, o).
\end{eqnarray*}
\noindent However, she does not know his problem ingredients $p_{o(x)}^{A}$, $u_{A}$. Suppose that we model her uncertainty about them through a random expectation $P_{o(x)}^{A}$ and random utilities $U_A(y_{C},y,o)$. Then, the random optimal obfuscation, when the malware features are $x$, will be
\begin{eqnarray*}
O^*(x,M) &=& \argmax_{o} \bigg( U_A( M, M, o) - U_A( B, M, o) \bigg) P_{o(x)}^A + U_A( B, M, o),
\end{eqnarray*}
and we would make $p_C(o_{x \rightarrow x'}|x,M) =Pr(A^*(x,M)=o_{x \rightarrow x'})$. Typically, we would approximate the attack probability $p_C(o_{x \rightarrow x'}|x,M)$ through Monte Carlo. We focus before on assessing the elements $U_{A}$ and $P_{o}^{A}(x)$. 

In relation with $U_{A}$, recall first that, without loss of generality, we may associate utility $0$ with the worst consequence and utility $1$ with the best one, having the other consequences intermediate utilities, e.g. \cite{french2000statistical}. In this problem, the best consequence for the attacker is that the classifier accepts a malware as benign (he then has opportunities to pursue his business), whereas his worst consequence holds when she stops a malware as such (he has wasted effort in a lost opportunity). The consequences related with benign binaries are in between (and are actually irrelevant for the Attacker's risk analysis). Therefore, we may actually say that $U_A(M,M,o) \sim \delta_{0}$ and $U_A(B,M,o) \sim \delta_{1}$, the degenerate distributions at $0$ and $1$, respectively.
\noindent Then, Alan's random optimal attack would be
\begin{eqnarray*}
O^*(x,M) &=& \argmax_{o} \bigg[\Big( 0 - 1 \Big) P_{o(x)}^A + 1 \bigg] = \argmax_{o} \bigg[ 1 - P_{o(x)}^A \bigg].
\end{eqnarray*}
    
As far as $P_{o(x)}^{A}$ is concerned, its assessment could be based on an estimate $r$ of $Pr_{C}(c(x') =M|x')$. As a probability, $r$ ranges in $[0,1]$ and we could make $P_{o(x)}^A \sim \beta e (\delta _1, \delta _2 )$, with mean $\delta _1 / (\delta _1 + \delta _ 2) = r$ and variance $ (\delta _1 \delta _2) / [(\delta _1 + \delta _2 )^2 (\delta _1 + \delta _2 + 1) ]=var$ as perceived, from which we obtain the parameters $\delta_{1}$ and $\delta_{2}$.  In order to estimate $r$, given a binary with features $x'$, we consider all reasonable attacks leading to it. Let $p_1$ be the malware probability estimates of these attacks; $p_2$, the benign probability estimates of these attacks. Then, we estimate $r$ through $r=p_1/(p_1+p_2)$. Note that such probability estimates are available from the training stage of the classifier.

We then use simulation with $L$ samples from the random probabilities, and find
\begin{eqnarray*}
O^*_l (x, M)=\argmax_{o}
\bigg[ 1 - P_{o(x)}^{A,l} \bigg], l=1,..,L
\end{eqnarray*}
estimating the required probability through
\begin{eqnarray}
\widehat{p _C} ( o_{x \rightarrow x'}\,|\,x, M) =  \frac {\# \{O_l^*(x, M) =  o_{x \rightarrow x'}\}} {L}. \label{mc_estimate}
\end{eqnarray}
%%%%%%%%%%%%%%%%%%%%%%%%%%%%
\subsection{An updated framework}
The hybrid framework for malware detection proposed in Section \ref{sec:framework} included an operational phase in which detection was based on standard risk analytic computations as in (\ref{eq:operation}). However, this is prone to be fooled by intelligent adversaries and we need to update the framework replacing it by the adversarial risk analytic version (\ref{pis}) which takes into account that the attacker might obfuscate the malware. At this point, we would stress 
also the need to replace the standard $0-1$ utility model used in classification by utilities better reflecting the severity of malware impact as illustrated in Section \ref{subsec:otherUtilites}.
%The proposed hybrid framework for malware detection comprised the stages of preprocessing, feature extraction, feature management, training and operation. In this last stage, we have already chosen the features and trained our classifier. However, we update the process, as in standard risk analysis through (\ref{eq:operation}) entailing that the framework in unaware of the attacker presence. Now, we consider the model proposed based on adversarial risk analysis that study the attacker's problem taking into account his preferences, his beliefs about the obfuscation attacks that he performs and his thoughts over the classifier predictions. At this point, the model is updated through (\ref{pis}) in which the classifier aims at obtaining the class $c(x')$ that maximises her expected utility given $x'$.

%%%%%%%%%%%%%%%%%%%%%%%%%%%%%%%%%%%%%%%%%%%%%
\section{Example}
\label{sec:examples}
We test the proposed approach using the developments in Sections \ref{sec:impactObfuscatedMalware} and \ref{sec:aroa}. The dataset is divided into training and test sets with a $0.20$ division ratio. We first train a Bernoulli NB with benign and non-obfuscated malware data. We perform $10$ groups of $100$ experiments averaging the results. We report the best results achieved based on the following parameters: Monte Carlo size $L=700$ in (3); variance $var=0.25$ for the 
$\beta e (\delta _1, \delta _2 )$ distribution; $300$ original binaries $x'$ generated leading to $x_{j}^{'}$ and $20$ attacks $o(x)$ leading to $x'$. 
To reach such settings, we performed an large quantity of experiments exploring 
ranges for the parameters; for instance, we tried the values $L \in \{100,200,\dots,1000\}$ in combination with the other parameters, achieving best results with $L=700$.

%%%%%%%%%%%%%%%%%%%%%%%%%
\subsection{$0-1$ Utility}
In the first set of experiments, we use the standard $0-1$ utility ($1$ if binary is correctly classified and $0$, otherwise) as presented in Table \ref{tab:uti0_1}, which leads to the standard DA classification criteria. Figure \ref{fig:acuUti0_1} represents the performance comparison of the proposed approach, which we designate Adversarial Risk Analysis for Obfuscation Attacks (AROA), with that of NB. 

\begin{minipage}{\textwidth}
%\centering
\begin{minipage}[t]{0.49\textwidth}
\centering
\vspace{35pt}
\begin{tabular}{c|c|c|c|}
\cline{3-4}
\multicolumn{1}{c}{} & \multicolumn{1}{c|}{}& \multicolumn{2}{c|}{\bf Actual}\\
\cline{3-4}
\multicolumn{2}{c|}{\bf } & $y = M$ & $y = B$ \\
\cline{1-4}
\multicolumn{1}{|c|}{\bf Predic.} & $y_{C} = M$ & 1 & 0 \\
\cline{2-4}
\multicolumn{1}{|c|}{\bf } & $y_{C} = B$ & 0 & 1 \\
\hline
\end{tabular}
\captionof{table}{$0-1$ utility\\ }\label{tab:uti0_1}
\end{minipage}
%\hfill
\begin{minipage}[t]{0.49\textwidth}
\centering
\vspace{0pt}
\includegraphics[width=0.85\textwidth]{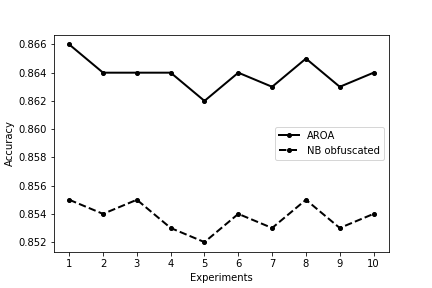}
\captionof{figure}{\small AROA vs NB, \\against obfuscated malware}\label{fig:acuUti0_1}
\end{minipage}
\end{minipage}
\vspace{20px}

\noindent Observe that AROA obtains higher accuracy than the best version of NB.

%%%%%%%%%%%%%%%%%%%%%%%%%%%%%%%%%%%%%%%%%%%%
\subsection{A non-standard utility function}
\label{subsec:otherUtilites}
We test now our AROA approach with other utilities,
penalising more false negatives (a malware binary classified as benign),
with, e.g., $-5$, 
than false positives (when a benign binary is classified as malware),
as we consider those much more harmful. This utility function is reflected in Table \ref{tab:utino5_1} and accuracy results presented in Figure \ref{fig:utino5_1}.

\begin{figure}[H] 
\includegraphics[width=0.47\textwidth]{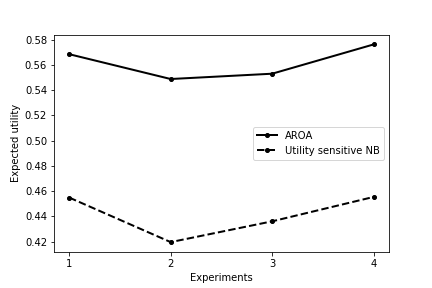}
\caption{utility sensitive NB vs AROA}
\label{fig:AROAvsUtNb}
\end{figure}

\begin{minipage}{\textwidth}
\begin{minipage}[t]{0.49\textwidth}
\centering
\vspace{35pt}
\begin{tabular}{c|c|c|c|}
\cline{3-4}
\multicolumn{1}{c}{} & \multicolumn{1}{c|}{}& \multicolumn{2}{c|}{\bf Actual}\\
\cline{3-4}
\multicolumn{2}{c|}{\bf } & $y = M$ & $y = B$ \\
\cline{1-4}
\multicolumn{1}{|c|}{\bf Predic.} & $y_{C} = M$ & 1 & 0 \\
\cline{2-4}
\multicolumn{1}{|c|}{\bf } & $y_{C} = B$ & -5 & 1 \\
\hline
\end{tabular}
\captionof{table}{Alternative utility}\label{tab:utino5_1}
\end{minipage}
\hfill
\begin{minipage}[t]{0.49\textwidth}
\centering
\vspace{0pt}
\includegraphics[width=0.85\textwidth]{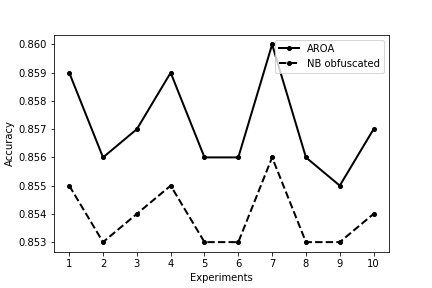}
\captionof{figure}{AROA vs NB,\\ against obfuscated malware}\label{fig:utino5_1}
\end{minipage}
\end{minipage}
\vspace{20px}
\noindent 
In this case, we again observe that AROA performs better than NB attaining bigger accuracy. 
More importantly, we have computed the associated expected utilities based on Table \ref{tab:utino5_1}. For this we used $4$ groups of $10$ experiments whose results are shown in Figure \ref{fig:AROAvsUtNb} which portray the larger expected utilities obtained with AROA
vs a standard (utility sensitive) NB approach.

%To end with the example, we would evaluate the expected utility using also the utilities from Table \ref{tab:utino5_1}. Figure \ref{fig:AROAvsUtNb} illustrates $4$ groups of $100$ experiments averaged, here we compare an utility sensitive NB with AROA approach.
%We note that AROA still acquiring higher expected utility when comparing with the utility sensitive NB.
Again we observe that AROA clearly outperforms NB, perhaps even more markedly, in terms of 
attained utility. Similar results have been attained with other utility functions.

%%%%%%%%%%%%%%%%%%%%%%%%%%%%%%%%%%%%%%%%%%%%%
\section{Discussion}
\label{sec:discussion}
Malware entails major cybersecurity risks as attackers learn to use advanced obfuscation techniques to degrade the detection capacities of standard algorithms. 
We showcased the potential degradation with a metamorphic tool used by attackers to obfuscate malware.
Due to these inefficiencies, we have proposed a novel approach based on adversarial risk analysis combined with a hybrid framework which extracts static and dynamic features from binaries.
 We analysed it observing which features are relevant during the obfuscation process and how their entailed data transformations affect NB performance. Our alternative approach based
 on adversarial risk analytic computations improve performance suggesting potential in applications, even more taking into account its operational speed.

There are many ways to continue this work. First, we have exemplified the approach
with NB classifiers; however many other algorithms have been used in this area as reviewed
in the Introduction and they could replace NB in our development. We have considered
a two-class problem (benign, malware) but there are other cybersecurity risk analysis 
problems which are multiclass (as with crime server classification) and we could extend the framework to such context. We have also considered attacks in the operational phase, but it could be the case that attacks take place in the training phase, and there is a need for further developments in this area as well.

\bibliographystyle{apalike}
\bibliography{references}

\begin{thebibliography}{}

\bibitem[{Ahuja}, 2017]{ahuja2017robust}
{Ahuja} (2017).
\newblock {\em Robust Malware Detection using Integrated Static and Dynamic
  Analysis}.
\newblock PhD thesis, Thesis, Indian Institute of Technology Kanpur.

\bibitem[Anderson et~al., 2012]{anderson2012improving}
Anderson, B., Storlie, C., and Lane, T. (2012).
\newblock Improving malware classification: bridging the static/dynamic gap.
\newblock In {\em Proceedings of the 5th ACM workshop on Security and
  artificial intelligence}, pages 3--14. ACM.

\bibitem[Ashari et~al., 2013]{ashari2013performance}
Ashari, A., Paryudi, I., and Tjoa, A.~M. (2013).
\newblock Performance comparison between na{\"\i}ve bayes, decision tree and
  k-nearest neighbor in searching alternative design in an energy simulation
  tool.
\newblock {\em International Journal of Advanced Computer Science and
  Applications (IJACSA)}, 4(11).

\bibitem[Banks et~al., 2015]{banks2015adversarial}
Banks, D.~L., Aliaga, J. M.~R., and Insua, D.~R. (2015).
\newblock {\em {Adversarial Risk Analysis}}.
\newblock Chapman and Hall/CRC.

\bibitem[Barrett, 2018]{barrett2018framework}
Barrett, M. (2018).
\newblock Framework for improving critical infrastructure cybersecurity.
\newblock {\em National Institute of Standards and Technology, Gaithersburg,
  MD, USA, Tech. Rep}.

\bibitem[Bedford et~al., 2001]{bedford2001probabilistic}
Bedford, T., Cooke, R., et~al. (2001).
\newblock {\em Probabilistic risk analysis: foundations and methods}.
\newblock Cambridge University Press.

\bibitem[{Chadwick}, 2017]{derekChadwick}
{Chadwick} (2017).
\newblock {Malware Detection}.

\bibitem[{Chronicle}, 2018]{virusTotal}
{Chronicle} (2018).
\newblock {Virus Total}.

\bibitem[Couce et~al., 2019]{couce2019}
Couce, A., Insua, D., and Kosgodagan, A. (2019).
\newblock {Assessing and forecasting cybersecurity impacts}.
\newblock Technical report, ICMAT-CSIC.

\bibitem[{Cuckoo Sandbox}, 2018]{cuckoo}
{Cuckoo Sandbox} (2018).
\newblock {Cuckoo Sandbox}.

\bibitem[Elingiusti et~al., 2018]{elingiusti2018malware}
Elingiusti, M., Aniello, L., Querzoni, L., and Baldoni, R. (2018).
\newblock Malware detection: A survey and taxonomy of current techniques.
\newblock In {\em Cyber Threat Intelligence}, pages 169--191. Springer.

\bibitem[{ENISA}, 2019]{enisa2019}
{ENISA} (2019).
\newblock {Threat Landscape Report 2018 15 Top Cyberthreats and Trends}.
\newblock Technical report, European Union Agency For Network and Information
  Security.

\bibitem[{FreeBSD}, 2019]{hexdump}
{FreeBSD} (2019).
\newblock {hexdump}.

\bibitem[French and Insua, 2000]{french2000statistical}
French, S. and Insua, D.~R. (2000).
\newblock {\em {Statistical Decision Theory}}.
\newblock Wiley.

\bibitem[{GNU Binutils}, 2019]{objectdump}
{GNU Binutils} (2019).
\newblock {objectdump}.

\bibitem[Joseph et~al., 2019]{joseph2019adversarial}
Joseph, A., Nelson, B., Rubinstein, B., and Tygar, J. (2019).
\newblock {\em Adversarial Machine Learning}.
\newblock Cambridge University Press.

\bibitem[Kadane and Larkey, 1982]{kadane1982subjective}
Kadane, J.~B. and Larkey, P.~D. (1982).
\newblock {Subjective probability and the theory of games}.
\newblock {\em Management Science}, 28:113--120.

\bibitem[Kakisim et~al., 2018]{kakisim2018analysis}
Kakisim, A.~G., Nar, M., Carkaci, N., and Sogukpinar, I. (2018).
\newblock Analysis and evaluation of dynamic feature-based malware detection
  methods.
\newblock In {\em International Conference on Security for Information
  Technology and Communications}, pages 247--258. Springer.

\bibitem[Kaushal et~al., 2012]{kaushal2012metamorphic}
Kaushal, K., Swadas, P., and Prajapati, N. (2012).
\newblock Metamorphic malware detection using statistical analysis.
\newblock {\em International Journal of Soft Computing and Engineering},
  2:49--53.

\bibitem[Lakhotia et~al., 2005]{lakhotia2005method}
Lakhotia, A., Kumar, E.~U., and Venable, M. (2005).
\newblock A method for detecting obfuscated calls in malicious binaries.
\newblock {\em IEEE Transactions on Software Engineering}, 31:955--968.

\bibitem[Lewis, 1998]{lewis1998naive}
Lewis, D.~D. (1998).
\newblock Naive (bayes) at forty: The independence assumption in information
  retrieval.
\newblock In {\em European conference on machine learning}, pages 4--15.
  Springer.

\bibitem[{Malwarebytes}, 2018]{malwareBytesReport2018}
{Malwarebytes} (2018).
\newblock {2019 State of Malware}.

\bibitem[Naveiro et~al., 2019]{naveiro2019adversarial}
Naveiro, R., Redondo, A., Insua, D.~R., and Ruggeri, F. (2019).
\newblock Adversarial classification: An adversarial risk analysis approach.
\newblock {\em International Journal of Approximate Reasoning}, 113:133--148.

\bibitem[Ortega, 2016]{metame}
Ortega, A. (2016).
\newblock {Metame}.

\bibitem[O’Kane et~al., 2016]{o2016detecting}
O’Kane, P., Sezer, S., and McLaughlin, K. (2016).
\newblock Detecting obfuscated malware using reduced opcode set and optimised
  runtime trace.
\newblock {\em Security Informatics}, 5(1):2.

\bibitem[Rad et~al., 2012]{rad2012camouflage}
Rad, B.~B., Masrom, M., and Ibrahim, S. (2012).
\newblock Camouflage in malware: from encryption to metamorphism.
\newblock {\em International Journal of Computer Science and Network Security},
  12:74--83.

\bibitem[Radai, 1992]{checksummingTechniquesForAntiviralPurposes1992}
Radai, Y. (1992).
\newblock Checksumming techniques for anti-viral purposes.
\newblock In {\em {Proceedings of the IFIP 12th World Computer Congress on
  Education and Society-Information Processing}}, pages 511--517. North-Holland
  Publishing Co.

\bibitem[Rao et~al., 2016]{rao2016defense}
Rao, N.~S., Poole, S.~W., Ma, C.~Y., He, F., Zhuang, J., and Yau, D.~K. (2016).
\newblock {Defense of Cyber Infrastructures Against Cyber-physical Attacks
  Using Game-theoretic Models}.
\newblock {\em Risk Analysis}, 36:694--710.

\bibitem[Rolles, 2009]{rolles2009unpacking}
Rolles, R. (2009).
\newblock Unpacking virtualization obfuscators.
\newblock In {\em 3rd USENIX Workshop on Offensive Technologies}.

\bibitem[Sahay and Chaudhari, 2019]{sahay2019efficient}
Sahay, S.~K. and Chaudhari, M. (2019).
\newblock An efficient detection of malware by naive bayes classifier using
  gpgpu.
\newblock In {\em Advances in Computer Communication and Computational
  Sciences}, pages 255--262. Springer.

\bibitem[Santos et~al., 2009]{santos2009n}
Santos, I., Penya, Y.~K., Devesa, J., and Bringas, P.~G. (2009).
\newblock N-grams-based file signatures for malware detection.
\newblock {\em ICEIS}, 9:317--320.

\bibitem[Singh et~al., 2019]{singh2019comparison}
Singh, G., Kumar, B., Gaur, L., and Tyagi, A. (2019).
\newblock Comparison between multinomial and bernoulli na{\"\i}ve bayes for
  text classification.
\newblock In {\em 2019 International Conference on Automation, Computational
  and Technology Management}, pages 593--596. IEEE.

\bibitem[Van~Nhuong et~al., 2014]{van2014sssm}
Van~Nhuong, N., Nhi, V. T.~Y., Cam, N.~T., Phu, M.~X., and Tan, C.~D. (2014).
\newblock Sssm-semantic set and string matching based malware detection.
\newblock In {\em Computational Intelligence for Security and Defense
  Applications, 2014 Seventh IEEE Symposium on}, pages 1--6. IEEE.

\bibitem[Vorobeychik et~al., 2019]{vorobeychik2018adversarial}
Vorobeychik, Y., Kantarcioglu, M., and Brachman, R. (2019).
\newblock {\em Adversarial Machine Learning}.
\newblock Synthesis Lectures on Artificial Intelligence and Machine Learning.
  Morgan \& Claypool Publishers.

\bibitem[VxHeaven, 2013]{heavenvx}
VxHeaven (2013).
\newblock Malware collection.

\bibitem[Wang et~al., 2009]{wang2009detecting}
Wang, T.-Y., Wu, C.-H., and Hsieh, C.-C. (2009).
\newblock Detecting unknown malicious executables using portable executable
  headers.
\newblock In {\em 2009 Fifth International Joint Conference on INC, IMS and
  IDC}. IEEE.

\bibitem[{World Economic Forum}, 2017]{theGlobalRisksReport2017}
{World Economic Forum} ({2017}).
\newblock {The Global Risks Report 2017}.

\bibitem[{World Economic Forum}, 2018]{theGlobalRisksReport2018}
{World Economic Forum} (2018).
\newblock {The Global Risks Report 2018}.

\bibitem[{World Economic Forum}, 2019]{theGlobalRisksReport2019}
{World Economic Forum} (2019).
\newblock {The Global Risks Report 2019}.

\bibitem[Ye et~al., 2017]{ye2017survey}
Ye, Y., Li, T., Adjeroh, D., and Iyengar, S.~S. (2017).
\newblock A survey on malware detection using data mining techniques.
\newblock {\em ACM Computing Surveys}, 50(3):41.

\bibitem[Yilmaz et~al., 2019]{yilmaz2019locally}
Yilmaz, E., Al-Rubaie, M., and Chang, J.~M. (2019).
\newblock Locally differentially private naive bayes classification.
\newblock {\em arXiv preprint arXiv:1905.01039}.

\bibitem[You and Yim, 2010]{you2010malwareObfuscationTechniques}
You, I. and Yim, K. (2010).
\newblock Malware obfuscation techniques: A brief survey.
\newblock In {\em Broadband, Wireless Computing, Communication and
  Applications}, pages 297--300. IEEE.

\end{thebibliography}

%%%%%%%%%%%%%%%%%%%%%%%%%%%%%%%%%%%%%%%%%%%%%
%\section*{Acknowledgements}
%\label{sec:acknowledgements}
%The work has been performed under Project HPC\-EUROPA3 (INFRAIA-2016-1-730897), with the support of the EC Research Innovation Action under the H2020 Programme. The authors gratefully acknowledge the computer resources and technical support provided by KTH-PDC as well as Virus Total for providing  malware binaries for this study. We are also grateful to for the support from projects MTM2017-86875-C3-1-R AEI/ FEDER, RTC-2017-6593-7 AEI/FEDER and the AXA-ICMAT Chair in Adversarial Risk Analysis.

%%%%%%%%%%%%%%%%%%%%%%%%%%%%%%%%%%%%%%%%%%%%%

\end{document}